# Bulk Synthesis of Iron-based Superconductors

Athena S. Sefat [*]

*Materials Science & Technology Division, Oak Ridge National Laboratory,
Oak Ridge, Tennessee 37831, USA*

[*] e-mail: sefata@ornl.gov
Phone: (865)574-5495
Fax: (865) 576-5023

**Abstract**

Exploratory synthesis efforts for iron-based superconductors (FeSC) have been driven by hopes of improving superconducting critical temperatures ($T_C$s), providing high-quality samples for in-depth studies of intrinsic properties, and exploring potential superconductivity in similar families of materials. This manuscript summarizes the synthesis routes that are used for producing FeSC and their undoped parents, in single crystal and polycrystalline forms. A few of the materials challenges are summarized.

## 1. Introduction

High-temperature superconductivity (HTS) is among the most mysterious and elusive properties in condensed matter physics. The design of new and improved superconducting materials is of crucial importance both for understanding the fundamental nature of the superconducting state and for fulfilling the promise of superconductive materials for widespread use in energy-related needs such as generators and transmission lines. The reasonably high $T_C$s, high upper critical fields ($H_{C2}$), relatively small anisotropy, and long coherence length make the FeSC encouraging for high-field applications. However, better properties are still needed in order to make superconductors widespread. The accidental discovery of Fe-based superconductors in an oxyarsenide in 2008 [1], which followed that of Ni- and Fe-based oxyphosphides with the same crystal structure [2 - 4], instigated extensive research efforts for understanding the causes of HTS, finding new FeSC, and for hunting for HTS in similar structure-types without iron.

The FeSC have similarities to the other well-known cuprate family of high temperature superconductors in that they have layered quasi two-dimensional tetragonal structures, feature square planar iron sheets analogous to the copper sheets in the cuprates, and are near antiferromagnetism (AF). However, unlike the cuprate, in which the Cu-Cu distances are long (~ 3.85 Å) with AF resulting from local-moment superexchange interactions, the Fe-Fe distances are short (2.67 to 2.85 Å) in FeSC; Fe 3$d$ states dominate the density of states at the Fermi level, and AF is a result of itinerant electrons nesting at the Fermi surface [5]. As a result, unlike the cuprates, in which the parents are Mott-Hubbard insulators, the parents of FeSC are spin-density-wave [6], with a Fermi surface that is very sensitive to small changes in chemical substitutions or pressure [5, 7]. Moreover, superconductivity can be caused by electron doping in Fe planes [8, 9], as opposed to the cuprates, in which the Cu layers need to stay intact. This may suggest that the role of Fe in FeSC may be less critical.

With the discovery of FeSC, and the extraordinarily rich crystal chemistry of this system, the synthesis of superconducting materials has expanded to more diversified techniques. The synthesis of FeSC is more difficult compared to that of the cuprates, due to the high toxicity of arsenic and selenium elements, the high vapor pressure of arsenic (at ~600 °C, 1 atm), the high moisture/oxygen reactivity and the low melting points of alkali metals and alkaline-earth metals, and the oxygen sensitivity of rare-earth metals. As a result, the preparation of FeSC is carried out in an inert-gas-filled glovebox, with ppm pressure levels of $O_2$ and $H_2O$. In addition, the materials are also air sensitive during reaction temperatures, so unlike the cuprates, the synthesis is complicated by the need to seal the reactions in vacuum or inert atmospheres. For single-crystal synthesis, although float-zone technique is primarily used for cuprates, flux-growth and Bridgman techniques are used for FeSC, as will be described below. For film deposition, physical vapor deposition and solution methods have been commonly employed for cuprates, such as pulsed injection metal organic chemical vapor deposition (MOCVD) and metal organic deposition (MOD) of precursors followed by decomposition and reaction anneals; however, deposition techniques involving high vacuum such as pulsed laser deposition (PLD) and molecular beam epitaxy (MBE) have been used for FeSC [10, 11].

All of the FeSC (> 50 compounds) have been discovered using chemical intuition and synthesis attempts. Such explosive growth of materials, along with experimental knowledge, has yet to be matched by theory. These materials are all based on tetragonal crystal structures that are made of FeAs or FeSe layers, with nominally divalent iron in tetrahedral coordination. FeSC can be categorized by chemical formula and structure type into the following five family types: '1111' ZrCuSiAs-type of $R$FeAsO ($R$ = rare earth) [1], '122' $ThCr_2Si_2$-type of $B$Fe$_2$As$_2$ ($B$ = alkaline earth) [12] or $B_{1-x}Fe_2Se_2$ ($B$ = alkali or alkaline-earth metal) [13], '111' $Cu_2Sb$-type of LiFeAs [14], '11' PbO-type of FeSe [15], and '42622' $Sr_2FeO_3CuS$-type of $Sr_4Sc_2O_6Fe_2P_2$ [16]. At room temperature, all of the families are in the $P4/nmm$ space group, with the exception of 122 in the $I4/mmm$. All of the families have similar room temperature $a$-lattice parameters of ≈ 3.9 Å, comparable to cuprates. The $c$-lattice parameters vary from ≈ 5.1 Å for 11, to 15.8 Å for 42622.

In FeSC, although the maximum $T_C$ to date is only 55 K, the materials are fairly isotropic, tolerate disorder well, show great chemical flexibility, and are amenable to first-principles approaches that should add fundamental knowledge to the causes of unconventional superconductivity. The maximum $T_C$ is found in the 1111 family by electron doping within the $R$O planes in $Gd_{0.8}Th_{0.2}FeAsO$ [17], $SmFeAsO_{0.9}F_{0.1}$ [18], $SmFeAsO_{0.85}$ [19], $SmFeAsO_{0.8}H_{0.2}$ [20], and $Sr_{0.5}Sm_{0.5}FeAsF$ [21]. In 122, $T_C$ as high as 47 K are found in arsenides by electron doping within $B$ sheets in $Ca_{0.8}La_{0.2}Fe_2As_2$ [22] and $Ca_{0.85}Pr_{0.15}Fe_2As_2$ [23], and in selenides by having off-stoichiometry in $Na_{0.7}Fe_{1.8}Se_2$ [24]. In 11, isovalent substitution gives highest $T_C$ = 15

K in $FeSe_{0.5}Te_{0.5}$ [25]. Slightly off-stoichiometric 111 and 42622 give $T_C$ = 18 K in $Li_xFeAs$ [14, 26] and $T_C$ = 37 K in '$Sr_4V_2O_6Fe_2As_2$' [27, 28].

Below is a brief survey of the literature's synthesis-strategic routes for making the five families of FeSC; for the complete list of all FeSC, several of which will not be discussed below, the reader is to refer to literature. The survey highlights the role of chemistry in the creation of all of the FeSC, illustrating the power of chemical intuition and exploratory synthesis efforts.

## 2. Overview of Synthesis Routes

In order to achieve a precise, correct, and deep insight into the causes of high-temperature superconductivity, good crystalline quality and non-amorphous materials are highly desirable. In a non-amorphous sample, the constituent atoms are arranged in an orderly pattern persisting over distances much larger than the size of the unit cell. A polycrystalline material can be defined as possessing ordered regions (grains) of ~ 1 - 10 μm (< $10^5$ Å) in size, and single crystals may be defined as having ~ 0.1 mm to cm (> $10^6$ Å) size crystallites. Whether the material is obtained as a single crystal or in polycrystalline form depends on the synthesis conditions used. Although the preparation of quality single crystals is crucial for finding the anisotropic and intrinsic properties, in some materials crystal preparation is very difficult. Although a crystal can be synthesized in a pristine form, small departures from regularity may exist in the forms of small imperfections and impurity inclusions. These defects can result in variations in composition and structure, leading to deviation from the intrinsic bulk transport and magnetic properties of the crystal.

The $T_C$ seems to be mainly determined by chemical composition and structure, whereas $H_{C2}$ and current density ($J_C$) are also influenced by microstructures that can be optimized by varying processing conditions. In order to understand the causes of superconductivity at a certain $T_C$, it is essential to study homogenous and high-purity single crystals with minimal strain and no grain boundaries. Only once the materials problems are overcome can the universal structural and property features among the different superconducting classes in cuprates and FeSC be determined.

For the FeSC and their parents, chemical reactions are not performed under ambient conditions due to the air sensitivity of the reactants (rare-earths, alkali metals, alkaline-earth metals), reactant toxicity and volatility at reaction temperatures (arsenic, selenium), and the desire to form products without oxygen (e.g., in 122, 111, 11) and to have the metal in low nominal oxidation state ($Fe^{2+}$). Thus in the preparation of FeSC, the reactions are often loaded into silica ampoules, evacuated (and sometimes backfilled with inert gas), and then sealed by melting the silica using a blow torch. In sealed tubes, however, significant pressures can build up, and care should be taken to avoid minor explosions.

Below, the five synthetic methodologies that are used for making the five families of FeSC are highlighted. A description of each synthesis method is given, along with the list of families that are prepared. It should be noted that the following is not an exhaustive review of synthesis conditions presented in the literature; further details can be found within the reference lists and additional search of literature.

## 2.1 Solid-state method

This is the most commonly used method for the preparation of a material with multi-elements, especially if no prior knowledge of synthesis conditions exists. This method involves elevated and prolonged heating of the correct molar proportions of the appropriate solid reactants (elements or binaries) in powder forms. Non-pure, hygroscopic, non-stoichiometric, and volatile reactants should be avoided, as impurities are impossible to remove at the end of the reaction. This method enables the synthesis of mainly thermodynamically stable phases, since the reaction occurs in the solid state (no melting) and requires the diffusion of ions across the grains' points of contacts. New structures (the desired material) can form at the phase boundary between the grains. In order to speed up the reactions in the solid-state method, the temperature can be increased to improve diffusion; the mixture can be ground to improve homogeneity and reduce particle size, and the powder mixture can be pelletized to improve grain contacts and to lessen voids.

The solid-state method has been used for the preparation of all of the five families of FeSC, with multiple steps, high sintering temperatures, and sealed reactions. Resistive-heating box furnaces and high melting-point silica ($T_{max}$~1250 °C) vessels have been used to contain the reactions, possibly in conjunction with alumina or precious metal (Au, Pt) containers. Because of the air-sensitivity or toxicity of many of the reactants, the air exposure is minimal or avoided.

For producing 1111 FeSC, fluorine can be chemically substituted for oxygen in $LaFeAsO_{1-x}F_x$ (max $T_C \approx 26$ K). The stoichiometric molar reactant mixtures are pelletized and sealed in silica tubes, and then heated at 1250 °C for ~ 40 hours. The discovery report on $LaFeAsO_{1-x}F_x$ uses the set of reactants listed in equation *a*; the x-ray diffraction pattern for x = 0.05 gives evidence of minority FeAs and LaOF crystalline impurities [1]. Because the grain surface of lanthanum powders is usually oxidized and lanthanum chunks cannot be ground and homogenized with other reactants, a proceeding report offers a different set of reactants that are listed in equation *b*, with LaAs prepared as a binary first [29]. In this report, ~5% impurity phase of $La_{4.67}(SiO_4)_3O$ phase is detected in x-ray diffraction of x= 0.11 ($T_C$ = 27 K), due to surface reaction of the pellet with the silica vessel containing the pellet because of the high anneal temperatures (1250 °C) and durations (40 hrs) [29]. A third report prepares this material ($T_C$ = 20 K) by using $Fe_2O_3$ as the source of oxygen (see equation *c*); the reactants are reacted longer and at a lower temperature of 1150 °C in a tantalum crucible that are sealed in silica [30]. Other fluorine-doped 1111 polycrystalline superconductors, with $T_C$ values as high as 55 K, are found by replacement of La with other rare-earths in similar reactions, initially in $SmFeAsO_{1-x}F_x$ and $NdFeAsO_{1-x}F_x$ [18, 31, 32].

(*a*) $(1-x)/3\ La_2O_3 + x/3\ LaF_3 + (1+x)/3\ La + FeAs \rightarrow LaFeAsO_{1-x}F_x$
(*b*) $(1+x)/3\ LaAs + ½\ Fe_2As + (1-x)/3\ La_2O_3 + x/3\ LaF_3 + (1-2x)/6\ As \rightarrow LaFeAsO_{1-x}F_x$
(*c*) $(1-x/3)\ LaAs + x/3\ LaF_3 + (1-x)/3\ Fe_2O_3 + (1 + 2x)/3\ Fe + x/3\ As \rightarrow LaFeAsO_{1-x}F_x$
(*d*) $LaAs + x/3\ Co_3O_4 + (3-4x)/9\ Fe_2O_3 + (3-x)/9\ Fe \rightarrow LaFe_{1-x}Co_xAsO$

For producing 1111 FeSC, cobalt can also be chemically substituted for iron in $LaFe_{1-x}Co_xAsO$ (max $T_C \approx 15$ K). For making this transition-metal doped 1111 superconductor, reactants in equation *d* are used, with several regrinding and annealing procedures at 1220 °C for 12 hrs; minor impurity phases of FeAs, $La_2O_3$ and $La(OH)_3$ are however reported [9].

For producing 122 FeSC, potassium can be chemically substituted for barium in $Ba_{1-x}K_xFe_2As_2$ (max $T_C \approx 38$ K). The stoichiometric molar amounts of the elements, listed in equation *e*, are placed inside an alumina crucible, sealed in a silica tube in partial pressure of argon, and then heated at 600 °C for 15 hours [12]. After the reaction is cooled, the products are ground and pressed into a pellet; annealing and regrinding procedures are then followed at 650 °C and 750 °C; minor impurity phase of FeAs (~ 6%) is reported [12]. For producing other types of 122 FeSC, phosphorus can isovalently substitute for arsenic in $BaFe_2(As_{1-x}P_x)_2$ (max $T_C \approx 32$ K); the list of reactants in equation *f* is used, and annealing and regrinding steps at 300 °C, 900 °C, and 1000 °C; minor secondary phase of $Fe_2P$ is detected [33].

(*e*) $(1-x)Ba + x K + Fe + As \rightarrow Ba_{1-x}K_xFe_2As_2$
(*f*) $Ba + 2Fe + 2(1-x)As + 2xP \rightarrow BaFe_2(As_{1-x}P_x)_2$

For producing 11 FeSC of $FeSe_x$ (max $T_C = 8$ K), stoichiometric amounts of selenium and iron powders are mixed according to equation *g*, ground, pelletized, and then annealed several times in evacuated silica tubes up to 700 °C [15]. Minor impurity phases of hexagonal $Fe_7Se_8$, Fe, FeSi, and $Fe_2O_3$ can be detected. Higher $T_C = 15$ K is produced by isovalent substitution of selenium for tellurium in $FeTe_xSe_{1-x}$ [25]. Using equation *h*, and annealing steps at 600 °C and 650 °C produces mainly pure products, with minor hexagonal $Fe_7Se_8$ impurity.

(*g*) $Fe + x Se \rightarrow FeSe_x$
(*h*) $Fe + (1-x)Se + xTe \rightarrow FeTe_xSe_{1-x}$

For producing 111 FeSC (max $T_C = 18$ K), highly air sensitive $Li_xFeAs$ is synthesized from elements according to equation *i*, or by using Li and FeAs binary [14, 26, 34]. The stoichiometric amounts of reactants are reacted up to 800 °C for ~ 1 day, in niobium [26] or tantalum tubes [34] that are backfilled with ~ 1 atm of argon, sealed in silica; FeAs impurity is noted in the products. 111 can also be synthesized in gold crucibles under high pressures of 1.8 GPa, at a reduced reaction time (~1 hr) and Li evaporation [26].

Superconductivity is also produced by chemical substitution of iron with cobalt (or nickel) in $NaFe_{1-x}Co_xAs$ (max $T_C = 20$ K) [35]; the elemental reagents (equation *j*) are sealed under 1 atm of argon in a niobium tube, and followed by annealing steps at 200 °C and 750 °C [35].

(*i*) $xLi + Fe + As \rightarrow Li_xFeAs$
(*j*) $Na + (1-x)Fe + xCo + As \rightarrow NaFe_{1-x}Co_xAs$

For producing 42622 FeSC (max $T_C = 37$ K), stoichiometric amounts of SrAs, $V_2O_5$, SrO, Fe and Sr can be ground and pressed into a pellet, followed by heating in sealed silica tube at 1150 ºC [27]. Another publication uses a different set of reactants namely $V_2O_5$, $SrO_2$, Sr, and FeAs, with heating at 750 ºC and then at 1150 ºC [36]. A more viable set of reactants is listed in equation *k*, since SrAs is hard to purity and chunks of Sr cannot be homogenized with other powders; highest $T_C$ is produced by heating at 850 ºC, then at 1150 ºC (24 hrs); products contained FeAs and $Sr_2VO_4$ impurities [28].

(*k*)  $4SrO + 2VAs + 2/3\ Fe + 2/3Fe_2O_3 \rightarrow Sr_4V_2O_6Fe_2As_2$

## 2.2 Bridgman method

In this synthesis method, congruently melting mixtures of elements or binaries are heated to slightly above their melting point in a sealed inert crucible, and crystals are grown by slow cooling from the melt. There exists a temperature gradient across the reaction such that crystals nucleate in the colder part of the crucible, growing slowly out of the melt. This method cannot be used to grow materials that have high melting points as containers and tubes for containing the reaction may be hard to find; also there may be difficulties in attaining and controlling the high temperatures required for the melt. Among the FeSC, 42622, 1111, and 111 families either have very high melting temperatures or decompose before melting.

The 11 family of $FeTe_xSe_{1-x}$ is readily grown via directional crystallization from the melt using this method, however, there are compositional gradients (x variations) and regions of phase segregation within a 'single crystal' boule [37]. In order to grow 11, molar reactant amounts of the reactants (equation *h*) are loaded into a tip-shaped silica ampoule, evacuated and sealed. Melting is ensured at 1070 ºC or lower, and the reaction is homogenized by dwelling for many hours (≥ 36 hrs), and then cooled in a temperature gradient furnace at ~4 ºC/hr [37]. Due to the thermal shock that the Bridgman silica vessel may experience, it is sealed into a second silica ampoule in order to hold vacuum during crystal growth. The Bridgman grown crystal boules can be more than a cm on a side; crystals can be easily cleaved perpendicular to the *c* axis (Figure 1).

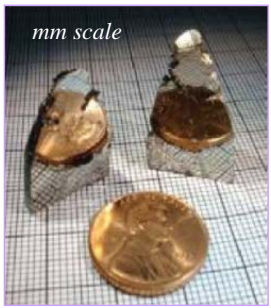

Figure 1: Bridgman-grown 11 iron-based superconductor, with average composition of $Fe_{\delta}Te_{0.75}Se_{0.25}$ ($T_C \approx 15$ K).

There is a report of 122 $BaFe_2(As_{0.68}P_{0.32})_2$ crystal (less than 1 mm size) grown using the Bridgman method [38]. For this, first a polycrystalline material is prepared similar to described above and equation *f*. The material is then placed into an alumina crucible and sealed inside a niobium container. An inductive furnace is used to reach temperatures well above 1250 ºC, then cooled (60 ºC/hr).

There are several reports for the synthesis of 122 selenides using Bridgman method, starting with $K_xFe_{2-y}Se_2$ ($T_C \approx 30$ K) giving mm size crystals [13]. However, alkali metal and iron contents along with the roles of chemical disorder and phase segregations, which seem to be important for obtaining superconductivity, remain to be matters of debate in these crystals.

## 2.3 Flux method

This method is used for growing crystals by slow-cooling a supersaturated solution. Materials that melt incongruently or at very high-temperatures, or that decompose before melting, can be grown out of a flux. The use of a flux allows for short crystal-growth time scales and significantly-lowered reaction temperatures. A flux is a low melting-temperature solvent that is used to dissolve reactants; the flux can be composed of metal elements (e.g., Zn, Ga, In, Sn), eutectic salts (e.g. NaCl/KCl), or binaries (e.g., FeAs). A good flux offers reasonable solubility and diffusivity for the reactants, and can be easily separated from the products; it does not enter the crystal as inclusions or chemical substitutions, nor does it form competing stable compounds with reactants. The flux method is commonly used for the crystal preparations of 122, and rarely for 11 and 1111, families of FeSC.

Solution growth from salts generally produces very small crystals, but the removal of salt is easy by rinsing in deionized water. The crystals of 1111 and 11 have been grown out of salt. $LaFeAsO_{1-x}F_x$ and $LaFe_{1-x}Co_xAsO$, as large as a few mm on a side, are grown out of NaAs ($T_{melt}$ = 600 ºC) [39]. For this, a solution with molar ratio of polycrystalline 1111:NaAs = 1:20 are sealed in a tantalum tube under partial argon atmosphere, and then sealed in an evacuated silica tube; the reaction is heated to 1150 ºC (dwell 24 hrs) and cooled to 600 ºC (3 ºC/hr). In order to ensure sufficient fluorine doping in $LaFeAsO_{1-x}F_x$, a small amount of NaF was added to NaAs, however, only partial maximum $T_C$ = 12 K is achieved; it is also reported that ~ 0.2 atomic % of Ta from the crucible material is incorporated inside these crystals [39]. $FeSe_x$ crystals, < 1 mm on a side, are grown out of excess KCl ($T_{melt}$ = 770 ºC) [40] or NaCl/KCl eutectic ($T_{melt}$ = 650 ºC) [41], in an evacuated sealed silica tube by heating up to 850 ºC, and slowly cooling to the solidification temperature of the salts.

Flux method from elemental metals of tin ($T_{melt}$ = 232 ºC) and indium ($T_{melt}$ = 157 ºC) are reported to produce mm size 122 crystals [40, 41]; however, metals get incorporated inside the crystals (up to 1 atomic %), and can dramatically change the intrinsic properties [42]. For producing $Ba_{1-x}K_xFe_2As_2$ crystals, the molar ratio of the elements and excess Sn are placed in an MgO crucible; the reaction is heated to 850 ºC, then slow cooled to 500 ºC. Flux is removed by the use of a centrifuge and a 'catch crucible' (an inverted crucible, containing silica wool, which is sealed on top of the reaction crucible). In addition to the metal flux incorporating in the crystal, it is reported that K concentration can vary along the length of a crystal due to its high volatility [42].

Flux method from FeAs binary material that already makes up part of the FeSC ('self-flux'), is reported to produce the largest and highest quality 122 crystals [8]. Crystals are reported to grow as well-formed plates, with the *c*-crystallographic direction perpendicular to the planes, with flat faces that are easily cleavable (Figure 2). Although FeAs is not a typical flux, in that it melts at a relatively high 1042 ºC, it has become the solution for producing single phase and large 122 FeSCs, as evident from several hundred reports on understanding their properties. In order to produce the flux of FeAs, care should be taken as arsenic sublimes before melting; the recipe is to place molar ratios of Fe and As inside a sealed thick-walled silica ampoule, slowly heat (30 ºC/hr) to 700 ºC (dwell 6 hrs), then (60 ºC/hr) to 1065 ºC (dwell 10 hrs), followed by cooling. Anisotropic properties of the first clean $BaFe_2As_2$ parent along with $BaFe_{2-x}Co_xAs_2$ superconductor ($\Delta T_C$ = 0.6 K; full shielding; $T_C$ = 22 K) was reported by using FeAs flux [8].

In order to grow single crystals out of transition-metal arsenides (*T*As), equations *l* and *m* can be followed by using molar ratio of 1:4 or 1:5 of alkaline-earth metal (*A*) to *T*As. The reactions are typically placed in alumina crucibles, and then sealed under partial argon atmosphere inside a silica tube; they are heated to 1180 ºC (dwell > 5 hrs), cooled slowly (~ 3 ºC/hr), followed by decanting of flux by the use of a centrifuge, above the melting temperature of flux (~1090 ºC for FeAs) [8, 44].

(*l*) $A + 5\ TAs \rightarrow AT_2As_2 + 3\ TAs$    (*A* = Ca, Sr, Ba; *T* = Cr, Mn, Fe, Co, Ni, Mo, etc.)
(*m*) $A + (5-x)\ FeAs + x\ TAs \rightarrow AFe_{2-x}T_xAs_2 + 3\ TAs$

Using this recipe and small variations of it, many 122 $AT_2As_2$ [45-47] and transition-metal doped $AFe_{2-x}T_xAs_2$ 122 crystals have been produced [48-51].

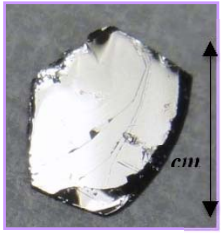

Figure 2: Typical size of the flux-grown 122 iron-based superconductor crystal; e.g., $BaFe_{2-x}Co_xAs_2$ ($T_C \approx 22$ K).

## 2.4 High-pressure method

This method offers the synthesis of metastable structures, extends the solubility limit of constituent elements, and produces materials with low oxidation states. The use of pressure also allows for considerably shortened reaction times and reduced evaporation of volatile elements. The main disadvantages of this method are that it is difficult to operate, small sample space results in small product yields, and small reaction times often result in unreacted reactants or side products.

This method is used for the preparations of 111 ($T_C$ = 18 K) [26] (described in 2.1), and of 1111 (up to $T_C$ = 56 K) that are oxygen-deficient, fluorine- or hydrogen-substituted. $RFeAsO_{1-y}$ (R = Nd, Sm, Gd, Tb, and Dy) are produced from pelletized stoichiometric mixtures of reactants that are loaded in a BN crucible, and then heated at ~ 1100 ºC under pressures of 2 to 5.5 GPa, for 2 hr [19, 52]. $PrFeAsO_{1-x}F_x$ is also produced from a stoichiometric mixture of well ground reactants, by sintering at 6 GPa and 1250 ºC for 2 hrs [53]. $SmFeAsO_{1-x}H_x$ [54] and $CeFeAsO_{1-x}H_x$ [55] are produced by reactions at 2 GPa and 1200 ºC.

## 2.5 Ammonothermal method

This synthesis method enables kinetic trapping of metastable products, which may be inaccessible or unstable at high synthesis-temperatures, by the use of the ammonia as solvent and catalyst, through redox chemistry. The synthesis of $A_{1-x}Fe_2Se_2$ FeSC (*A* = Li, Na, Ba, Sr, Ca, Yb, and Eu) is possible by this method and the use of Schlenk apparatus, producing superconducting shielding fractions up to 60% and $T_C$ values up to 46 K [56]. The synthesis process involves the transfer of an assembly containing FeSe and *A* to a gas-vacuum manifold, chilling to liquid nitrogen temperatures, opening to pure gaseous $NH_3$ atmosphere, which is then condensed in the chilled tube as solid $NH_3$. This solvent is then allowed to warm to the reaction, and finally removed by warming to room temperature and collect the gas for reuse in a chilled vessel. It is however reported that some ammonia always gets intercalated in the crystal structure [57, 58]; this may actually contribute to the stability of superconductivity and such high $T_C$ values in 122 selenides.

## 3. Conclusions

Ongoing synthesis efforts are important, as is evident from the serendipitous discovery of HTS in both cuprates and iron-based superconductors, then the preparations of their many families. Although the cause of HTS remains elusive, much effort continues to be applied to exploratory synthesis of new superconducting materials by seeking certain crystal structures, chemical substitutions, and promising properties; aside from the creation of potentially useful superconducting energy applications, such materials are essential for understanding HTS.

The design of a new superconductor involves predicting a suitable chemical composition to yield the desired structure and property, and the use of an appropriate synthesis method that would result in the needed bulk material. Both of these steps, however, are non-trivial, as not only are we unable to definitely predict the structure, stability, and properties of a superconductor from first principles, but the best method of synthesizing the material in pure single-crystal form is often difficult, especially when it involves many elements.

There is no doubt that if the exploration of further chemistry through classical synthesis and other sophisticated methods such as high-pressure synthesis and solvothermal method continues, new superconductivity discoveries will be made. The availability and knowledge of such a wide variety of synthetic strategies, combined with a better understanding of crystal chemistry, electronic structure, and thermodynamic and kinetic aspects of synthesis, will eventually enable the rational design of superconductors.

Because $T_C$ is mainly determined by chemical composition and crystal structure, it may even be that superconductivity in iron-based superconductors will eventually exceed that of the cuprates.

**Acknowledgement**

This work was supported by the Department of Energy, Basic Energy Sciences, Materials Sciences and Engineering Division.